\documentclass[11pt,a4paper]{article}
\usepackage{jheppub}
\usepackage{graphicx}
\usepackage{xcolor}
\usepackage{orcidlink} 
\usepackage[makeroom]{cancel}
\usepackage{enumitem}
\usepackage{amssymb,amsmath,mathrsfs,bm,bbm,mathtools} 
\usepackage{amsthm}
\definecolor{ao(english)}{rgb}{0.0, 0.5, 0.0}
\hypersetup{colorlinks=true,linkcolor=blue,urlcolor=blue,citecolor=ao(english)}
\usepackage{mathrsfs}
\usepackage[latin1,utf8]{inputenc}
\usepackage[makeroom]{cancel}

\newcommand\nn{\nonumber\\}

\newcommand{\bma}{\left(\begin{array}}
\newcommand{\ema}{\end{array}\right)}
\newcommand{\be}{\begin{equation}}
\newcommand{\ee}{\end{equation}}
\newcommand{\ben}{\begin{equation*}}
\newcommand{\een}{\end{equation*}}
\newcommand{\ba}{\begin{eqnarray}}
\newcommand{\ea}{\end{eqnarray}}
\newcommand{\ban}{\begin{eqnarray*}}
\newcommand{\ean}{\end{eqnarray*}}
\newcommand{\bs}{\begin{subequations}}
\newcommand{\es}{\end{subequations}}
\newcommand{\bc}{\begin{center}}
\newcommand{\ec}{\end{center}}

\newcommand{\Pl}{{\text{\tiny Pl}}}

\newcommand{\Mpl}{M_\Pl}

\newcommand{\au}[2]{#1.~#2}
\newcommand{\arX}[1]{\href{http://arxiv.org/abs/#1}{{\cob arXiv:#1}}}
\newcommand{\oarX}[1]{\href{http://arxiv.org/abs/#1}{{\cob #1}}}

\newcommand{\book}[5]{\emph{#1}, #2, #3, #4 (#5)}
\newcommand{\books}[4]{\emph{#1}, #2, #3 (#4)} 
\newcommand{\doin}[6]{\href{http://dx.doi.org/#1}{{\cob {\it #2 #3} {\bf #4} (#6) #5}}}
\newcommand{\doinn}[5]{\href{http://dx.doi.org/#1}{{\cob {\it #2} {\bf #3} (#5) #4}}}
\newcommand{\doij}[5]{\href{http://dx.doi.org/#1}{{\cob {\it #2} {\bf #3} (#5) #4}}}

\newcommand{\ndoinn}[5]{\href{#1}{{\cob {\it #2} {\bf #3} (#5) #4}}}

\newcommand{\procm}[6]{in \emph{#1}, #2 (eds.), #3, #4, #5 (#6)}
\newcommand{\tia}[1]{\textit{#1},}

\renewcommand{\leq}{\leqslant}
\renewcommand{\geq}{\geqslant}
\newcommand{\Eq}[1]{(\ref{#1})}
\newcommand{\Eqq}[1]{eq.~(\ref{#1})}
\newcommand{\Eqqs}[1]{eqs.~(\ref{#1})}
\def\rme{e}
\def\rmd{d}
\def\rmi{i}

\def\Re{\text{Re}}
\def\Im{\text{Im}}

\def\a{\alpha}
\def\b{\beta}
\def\de{\delta}

\def\g{\gamma}
\def\la{\lambda}
\def\k{\kappa}
\def\e{\epsilon}

\def\Om{\Omega}
\def\om{\omega}
\def\G{\Gamma}
\def\t{\tau}
\def\s{\sigma}

\def\vp{\varphi}
\def\N{\nabla}
\def\B{\Box}
\def\H{{\rm H}}

\def\Hes{{\bf\Delta}}

\def\cD{\mathcal{D}}

\def\cG{\mathcal{G}}
\def\cH{\mathcal{H}}

\def\cJ{\mathcal{J}}
\def\cK{\mathcal{K}}
\def\cL{\mathcal{L}}
\def\cM{\mathcal{M}}

\def\cO{\mathcal{O}}

\def\p{\partial}

\def\cob{\color{blue}}

\begin{document}

\title{Path integral and conformal instability in nonlocal quantum gravity} 

\author[a]{Gianluca Calcagni\,\orcidlink{0000-0003-2631-4588}}
\emailAdd{g.calcagni@csic.es}
\affiliation[a]{Instituto de Estructura de la Materia, CSIC, Serrano 121, 28006 Madrid, Spain}

\author[b]{and Leonardo Modesto\,\orcidlink{0000-0003-2783-8797}}
\emailAdd{leonardo.modesto@unica.it}
\affiliation[b]{Dipartimento di Fisica, Universit\`a di Cagliari, Cittadella Universitaria, 09042 Monserrato, Italy}

\abstract{We introduce the Lorentzian path integral of nonlocal quantum gravity. After introducing the functional measure, the Faddeev--Popov sector and the field correlators, we move to perturbation theory and describe Efimov analytic continuation of scattering amplitudes to Euclidean momenta and back to Lorentzian. We show that the conformal instability problem in the Euclidean path integral is solved by suitable gauge choices at the perturbative level. The three examples of Einstein gravity, Stelle gravity and nonlocal quantum gravity are given.}


\maketitle

\tableofcontents


\section{Introduction}

In the last five decades, the quest for a theory where the gravitational interaction and quantum mechanics can be combined consistently has seen a resurgence of candidates \cite{BMS}, to the point where, although we do not yet have a single fully consistent and controllable theory dominating the others, it is no longer tenable to claim that we do not know at all how to quantize gravity. If anything, the problem is that we are aware of far too many ways in which we could do it and too few ways to test their predictions with observations. Supergravity \cite{WeB,FvP}, the low-energy limit of string theory \cite{GSWi,Pol98,BBSb,Zwi09}, group field theory \cite{Ori09,BaO11,Fousp}, loop quantum gravity and spin foams \cite{Rov07,Per13,Ashtekar:2021kfp,Engle:2023qsu}, asymptotically safe quantum gravity \cite{Niedermaier:2006wt,Eichhorn:2018yfc,Bonanno:2020bil,Saueressig:2023irs}, causal dynamical triangulations \cite{Ambjorn:2012jv,Loll:2019rdj,Ambjorn:2024pyv} and others combine different notions of spacetime, quantization methods and dynamics. The degree of success in quantizing gravity and the level of completion of our understanding of the properties and phenomenology of the theory greatly vary among these scenarios. While some of them are nonperturbative, others are based on perturbative quantum field theory (QFT), let it be defined on a group manifold as in group field theory, on a higher-dimensional smooth spacetime where all fields enjoy supersymmetry as in supergravity, or on more minimalistic scenarios with a four-dimensional spacetime and without supersymmetry. In the latter case, of course, the price to pay to stay in a conservative perturbative QFT setting is to add new ingredients or to modify some of the traditional ones, such as making the dynamics fundamentally nonlocal (nonlocal quantum gravity, minimally \cite{Kra87,Kuzmin:1989sp,Tomboulis:1997gg,Modesto:2011kw,Biswas:2011ar,BCKM,Modesto:2014lga,Dona:2015tra,Koshelev:2016xqb,Modesto:2016max,CMN3,Modesto:2017sdr,Briscese:2018oyx,Buoninfante:2018mre,BasiBeneito:2022wux,Buoninfante:2022ild} or nonminimally \cite{Modesto:2021ief,Modesto:2021okr,Modesto:2021soh,Modesto:2022asj,Calcagni:2023goc} coupled to matter), imposing a prescription on the propagators and a projection on the spectrum to get rid of physical ghost modes (fakeon gravity \cite{	Anselmi:2017yux,Anselmi:2017lia,Anselmi:2018kgz,Anselmi:2018ibi,Anselmi:2018tmf,Anselmi:2018bra,Anselmi:2019rxg,Anselmi:2019nie,Anselmi:2021hab}), or giving up the notion of point particles in favour of a gas of quasi-particles living on a fractal spacetime (fractional gravity \cite{Calcagni:2016azd,Calcagni:2021ipd,Calcagni:2021aap,Calcagni:2022shb}).

Among the most recent perturbative approaches, the one that has perhaps received more attention is nonlocal quantum gravity or, more precisely, asymptotically local quantum gravity. The action of the theory is characterized by asymptotically polynomial nonlocal operators, entire functions of the Laplace--Beltrami operator $\B$ which do not add extra poles in the propagator while, at the same time, taking the form of finite-order polynomials in the ultraviolet (UV). Thanks to these properties, perturbative unitarity is preserved, a power-counting analysis of divergences is possible and there are robust indications that the theory is super-renormalizable or, in some of its versions, finite. Despite these advances, however, certain basic aspects of the theory have not been discussed with the due emphasis. For example, many works have been devoted to Feynman diagrams and scattering amplitudes but little has been said about their origin from a path integral. Moreover, these diagrams and amplitudes are invariably calculated in Euclidean momentum space due to the difficulty, or even impossibility, to handle a well-defined nonlocal QFT exclusively in Lorentzian signature. Therefore, questions may arise on whether the fundamental formulation of nonlocal quantum gravity is based on a Lorentzian or a Euclidean path integral, whether such path integral is convergent, and so on.

It is the purpose of this paper to address these questions. The theory is defined by a Lorentzian path integral, which is presented in detail in section \ref{sec2} together with its perturbative expansion and the quantum effective action. We will start from a purely gravitational action in $D$ topological dimensions:
\be
S = \frac{\Mpl^{2}}{2}\int\rmd^D x \sqrt{|g|}\,\cL\,,
\ee
where $\cL$ is the Lagrangian, we work in mostly plus signature $(-,+,\cdots,+)$ and $\Mpl\coloneqq (8\pi G)^{-1/2}$ is the reduced Planck mass with energy dimensionality $[\Mpl]=(D-2)/2$. When giving a concrete name to the Lagrangian, we will consider four theories:
\begin{itemize}
\item Einstein gravity:
\be
\cL=R\,,
\ee
where $R=g_{\mu\nu}R^{\mu\nu}=g_{\mu\nu}R_\s^{\mu\s\nu}$ is the Ricci scalar and $R^\rho_{~\mu\s\nu}\coloneqq \p_\s \G^\rho_{\mu\nu}-\p_\nu \G^\rho_{\mu\s}+\G^\tau_{\mu\nu}\G^\rho_{\s\tau}-\G^\tau_{\mu\s}\G^\rho_{\nu\tau}$ is the Riemann tensor.
\item Stelle gravity \cite{Stelle:1976gc,Stelle:1977ry,Julve:1978xn,Fradkin:1981hx,Fradkin:1981iu,Avramidi:1985ki,Hindawi:1995uk,Hindawi:1995an}:
\be
\cL=R+\g_0 R^2 + \g_2 R_{\mu \nu}R^{\mu \nu}+ \g_4 R_{\mu\nu\s\tau}R^{\mu\nu\s\tau},
\ee
where $\g_{0,2,4}$ are constants of dimensionality $[\g_i]=-2$.
\item Minimally coupled nonlocal quantum gravity \cite{Kra87,Kuzmin:1989sp,Tomboulis:1997gg,Modesto:2011kw,Biswas:2011ar,BCKM,Modesto:2014lga,Dona:2015tra,Koshelev:2016xqb,Modesto:2016max,CMN3,Modesto:2017sdr,Briscese:2018oyx,Buoninfante:2018mre,BasiBeneito:2022wux,Buoninfante:2022ild}:
\be\label{GA}
\cL=R + R \g_0(\B) R + R_{\mu\nu} \g_2(\B) R^{\mu\nu}+ R_{\mu\nu\s\tau}\g_4(\B)R^{\mu\nu\s\tau},
\ee
where the $\g_i(\B)$ are form factors depending on the Laplace--Beltrami operator $\B$ and, in principle, on three energy scales $\Lambda_{1,2,4}$ to make the arguments $\B/\Lambda_{0,2,4}^2$ dimensionless. The first two form factors can be parametrized as \cite{Dona:2015tra,Koshelev:2016xqb,Modesto:2016max}
\be
\g_0 =- \frac{(D-2)(\rme^{\H_0}-1) + D(\rme^{\H_2}-1)}{4(D-1)\B}+\g_4\,,\qquad
\g_2 =\frac{\rme^{\H_2}-1}{\B}-4\g_4 \,,\label{gamma}
\ee
where $\H_{0,2}(\B)$ are two entire functions which, in the case of asymptotically local quantum gravity, are asymptotically polynomial in the ultraviolet (UV) and such that $\H_{0,2}(0)=0$ in the infrared (IR). The reason for this specific form of nonlocal operators will be discussed in section \ref{sec3}. For simplicity, one can set $\H_0(\B)=\H_2(\B)=\H(\B)$ but we will not do that until later. Due to the Gauss--Bonnet theorem, the Riemann--Riemann term and its associated form factor $\g_4$ play no role in the renormalizability of the theory, since its presence amounts to a redefinition of the form factors appearing in the graviton propagator and scattering amplitudes:
\be\label{gprime}
\g_0\to\g_0'=\g_0-\g_4\,,\qquad \g_2\to\g_2'=\g_2+4\g_4\,.
\ee
\item Nonminimally coupled nonlocal quantum gravity \cite{Modesto:2021ief,Modesto:2021okr,Modesto:2021soh,Modesto:2022asj,Calcagni:2023goc}: in the absence of matter, 
\ba
&&\cL=G_{\mu\nu}F^{\mu\nu\s\t}(\B)G_{\s\t}\,,\qquad G_{\mu\nu}\coloneqq R_{\mu\nu}-\frac12 g_{\mu\nu}R\,,\\ &&F^{\mu\nu\s\t}(\B)=g^{\mu\nu}g^{\s\t}\frac{1}{(D-2)^2}\left[4\g_0(\B)-(D-4)\g_2(\B)\right]+g^{\mu\s}g^{\nu\t}\g_2(\B)\,,\qquad \g_4=0\,.\nn
\ea
Then, the action reduces to the same as in the minimally coupled case, with $\g_4=0$. The form factors in the class of nonlocal theories considered here are of asymptotically polynomial type,
\be
\rme^{\H_{0,2}(z)}= \rme^{\g_\textsc{e} + \G[0,p_{0,2}(z)]} \,  p_{0,2}(z)\,,\label{MCFormFactor}
\ee
where $\g_\textsc{e}$ is the Euler--Mascheroni constant, $\G$ is the upper incomplete gamma function and $p_{0,2}(z)$ are two polynomials of the same degree (this condition is required for renormalizability \cite{Modesto:2011kw,Modesto:2014lga}). In the IR, $\H_{0,2}(z)\simeq p_{0,2}(z)$ and $\exp[\H_{0,2}(z)]\simeq 1+p_{0,2}(z)$ if the polynomials do not have a constant term, while in the UV $\exp[\H_{0,2}(z)]\simeq\rme^{\g_\textsc{e}} p_{0,2}(z)$, hence the name asymptotically polynomial.
\end{itemize}

In section \ref{efi}, we recall the analytic continuation of scattering amplitudes to Euclidean signature. In section \ref{sec3}, we calculate the free-level (i.e., tree-level) graviton propagator for the generic theory \Eq{GA} in a general gauge, without specifying the form factors; the final result applies to all the above cases upon the choice of $\g_{0,2,4}$. In section \ref{secpi}, we discuss the conformal instability problem in the Euclidean path integral and show that it can be solved by gauge invariance at the \emph{perturbative} level for any of the above theories. In general, the conformal instability arises when one ignores gauge-fixing terms in the action, which are necessary to invert the graviton kinetic term. Section \ref{sec5} contains our conclusions and future extensions of these results, in particular, at the nonperturbative level.


\section{Lorentzian path integral}\label{sec2}

The path-integral quantization of nonlocal quantum gravity in the background-field gauge follows the same line as that of higher-derivative local theories \cite{BOS}. In this section, we consider only the gravitational sector of nonlocal quantum gravity, i.e., the action \Eq{GA}, without specifying the type of form factors (hence, what follows applies also to fractional gravity \cite{Calcagni:2016azd,Calcagni:2021ipd,Calcagni:2021aap,Calcagni:2022shb}). Introduce the background metric $\tilde{g}_{\mu\nu}$ and the Green's functions generating functional
\ba
Z[\tilde{g},J] \coloneqq \rme^{\rmi W[\tilde{g},J]}\coloneqq \int  \cD g_{\mu\nu} \cD \bar{C}_\alpha \cD C^\beta \cD b^\s 
\rme^{\rmi \left(S[g] + S_{\rm gf}[\tilde{g},g] + S_{\rm gh}[\tilde{g},g,C,\bar{C},b] + 
\int_{\tilde{g}} g_{\mu\nu} J^{\mu\nu}\right)},\label{Zgtilde}
\ea
where $S[g]$ is a generic gravitational action, while the action terms for the gauge fixing $S_{\rm gf}[\tilde{g},g]$, the Faddeev--Popov ghosts $C, \bar{C}$ and the third ghost $b$ are
\ba
S_{\rm gf}[\tilde{g},g] &=& \frac{1}{2} \int_{\tilde{g}} \chi_\alpha[\tilde{g},g] \, \cG^{\alpha \beta}[\tilde{g}] \, \chi_\beta[\tilde{g},g]
\,, \label{gaugefixing} \\
S_{\rm gh}[\tilde{g},g,C,\bar{C},b] &=& \int_{\tilde{g}} \bar{C}_\alpha M^\alpha_\beta[\tilde{g},g] C^\beta
+\frac{1}{2}  \int_{\tilde{g}}b_\alpha \cG^{\alpha \beta}[\tilde{g}] b_\beta\label{ghostsCb}\, ,
\ea
where we introduced the definition
\be
\int_{\tilde{g}}\coloneqq \int \rmd^D x \, \sqrt{|\tilde{g}|}
\ee
and the functionals $\cG^{\alpha \beta}[\tilde{g}]$ depending only on the background metric $\tilde{g}_{\mu\nu}$. The explicit form of the matrix $\cG^{\alpha \beta}[\tilde{g}]$ will be given later. The energy dimensionality of the elements in \Eq{gaugefixing} are $[\chi_\a]=1$ and $[\cG^{\alpha \beta}]=D-2$.

The integration measure we wrote in the simplified notation ``$\cD g_{\mu\nu} \cD \bar{C}_\alpha \cD C^\beta \cD b^\s$'' is actually more complicated and includes factors of the metric determinant preserving diffeomorphism invariance. In $D$ dimensions \cite{Fujikawa:1984qk,Anselmi:1991wb,Anselmi:1992hv,Anselmi:1993cu},
\ba
\cD g_{\mu\nu} \cD \bar{C}_\a \cD C^\b \cD b^\s &=& \prod_{\mu\leq\nu}\cD\left[|g|^\frac{D-4}{4D}g_{\mu\nu} \right]\,\prod_\a \cD\left[|\tilde g|^\frac{D-2}{4D}\bar C_\a\right]\nn
&&\times\prod_\b \cD\left[|g|^\frac{D+2}{4D}C^\b\right]\,\prod_\s \cD\left[|g|^\frac{D+2}{4D}b^\s\right], 
\ea
where the determinant associated with $\bar C_\a$ is of the background metric because the anti-ghost $\bar C_\a$ is a vector under background parametrizations.

Let us recall how the ghost modes in the generating functional \Eq{Zgtilde} emerge \cite{BOS}. Ignore the source $J$ for the time being. The generating functional for any metric gravitational theory and a given classical background $\tilde g_{\mu\nu}$ should be something of the form
\ben
Z[\tilde{g}] =\int_\cM \cD g_{\mu\nu}\,\rme^{\rmi S[g]}\,,
\een
where $\cM$ is the space of all possible metrics $g_{\mu\nu}=\tilde g_{\mu\nu}+h_{\mu\nu}$, $h_{\mu\nu}$ are quantum fluctuations of the metric and $S[g]$ is the classical action. However, this expression overcounts the metrics due to gauge invariance; in the case of gravity, diffeomorphism invariance. Functional integration should run only over physical metrics, i.e., metrics that are not equivalent under a diffeomorphism transformation over the manifold coordinates. Calling $\cG={\rm Diff}$ the diffeomorphism group of transformations on such manifold, the generating functional in terms of physical fields or gauge orbits $g_{\mu\nu}^{\rm phys}$ living in the quotient space $\cM/{\rm Diff}$ is
\ben
Z[\tilde{g}] =\det({\rm Diff})\int_{\cM/{\rm Diff}}\cD g_{\mu\nu}^{\rm phys}\,\rme^{\rmi S[g]}\,,
\een
where we factored out the gauge-group volume. This functional integral can also be expressed in terms of the original field $g$ via a constrained surface $\chi_\a[\tilde g,g]-l_\a=0$ in the space $\cM$, where $l_\a=l_\a(x)$ are arbitrary functions of spacetime coordinates $x$. Employing the Faddeev--Popov gauge-fixing procedure \cite{BOS,Sre07}, one can show that
\be\label{Zfapo}
Z[\tilde{g}] =\int_\cM\cD g_{\mu\nu}\,\rme^{\rmi S[g]}\de(\chi_\a-l_\a)\,\det M\,,
\ee
where $\chi_\a$ and the matrix $M^\a_\b$ depend both on the total metric $g_{\mu\nu}$ and on the background metric $\tilde g_{\mu\nu}$ separately. The gauge condition is expressed in terms of a functional $t_\alpha^{\mu\nu}[\tilde{g}]$ depending only on the background metric $\tilde{g}_{\mu\nu}$,
\be
\chi_\alpha[\tilde{g},g] = t_\alpha^{\mu\nu} [\tilde{g}] \, g_{\mu\nu}\,,\label{xi1}
\ee
which is linear in the field $g_{\mu\nu}$. The ghost operator $M^\alpha_\beta[\tilde{g},g]$ is
\be
M^\alpha_\beta[\tilde{g},g] = \frac{\delta \chi^\alpha}{\delta g_{\mu\nu}} \frac{\delta g_{\mu\nu}}{\delta \zeta^\beta}= t^{\alpha\mu\nu} [\tilde{g}]  \,r_{\mu\nu\beta} [g] \, , 
\label{Mfirst}
\ee
where the generators of the infinitesimal diffeomorphism transformations\index{diffeomorphism} $x^{\prime \alpha} = x^\alpha - \zeta^\alpha$ with parameter $\zeta^\alpha$ are defined by
\ba
&& \delta g_{\mu \nu} =  r_{\mu\nu\alpha} [g] \, \delta \zeta^\alpha \,, \nn
&& r_{\mu\nu\alpha} [g] \coloneqq g_{\mu\alpha} \partial_\nu + g_{\nu\alpha} \partial_\mu - \partial_\alpha g_{\mu\nu} 
= g_{\mu\alpha} \nabla_\nu + g_{\nu\alpha} \nabla_\mu\, .\label{infDiff}
\ea 

Given an arbitrary non-degenerate matrix $\cG^{\a\b}[\tilde g]$, one can use the identity
\be
\sqrt{\det\cG}\int\cD l_\a\,\exp\left(\frac{\rmi}{2}\int_{\tilde g} l_\a\cG^{\a\b}[\tilde g]\,l_\b\right)=\mathbbm{1}
\ee
to reexpress \Eqq{Zfapo} as
\be\label{Zfapo2}
Z[\tilde{g}] =\int_\cM \cD g_{\mu\nu}\,\exp\left(\rmi S[g]+\frac{\rmi}{2}\int_{\tilde g} \chi_\a\cG^{\a\b}\chi_\b\right)\det M\,\sqrt{\det\cG}\,.
\ee
The last two terms, also known as Faddeev--Popov determinants, can be written as functional integrals of the ghost spinors $C_\a$, $\bar C_\a$ and the ghost vector boson $b_\a$:
\bs\ba
\det M &=&\int\cD \bar{C}_\a \cD C^\b\,\exp\left(\rmi\int_{\tilde g} \bar C_\a M^\a_\b C^\b\right),\\
\sqrt{\det\cG} &=& \int\cD b_\a\,\exp\left(\frac{\rmi}{2}\int_{\tilde g} b_\a \cG^{\a\b}b_\b\right).
\ea\es
This leads to the final form \Eq{Zgtilde} of \Eqq{Zfapo2}.

Now we have all the ingredients to introduce the quantum effective action. 


\subsection{Quantum effective action}\label{qeasec}

Define the mean field
\be
\bar{g}_{\mu\nu} \coloneqq \frac{1}{\sqrt{|\tilde{g}|}} \frac{\delta W[\tilde{g},J]}{\delta J^{\mu\nu}} 
\label{MeanField}
\ee
and the quantum effective action
\be
\Gamma[\tilde{g},\bar{g}] \coloneqq W[\tilde{g},J] - \int_{\tilde{g}} \bar{g}_{\mu\nu} J^{\mu\nu} .
\label{QEAG}
\ee
Equations \Eq{MeanField} and \Eq{QEAG} lead to the equation of motion for the mean field
\be
\frac{1}{\sqrt{|\tilde{g}|}} \frac{\delta \Gamma[\tilde{g},\bar{g}]}{\delta \bar{g}_{\mu\nu}} = -  J^{\mu\nu} \, .
\label{extJ}
\ee 

Making the field redefinition
\be
g_{\mu\nu} =  \bar{g}_{\mu\nu} + h_{\mu\nu} 
\ee
in the functional integral (\ref{Zgtilde}) and using the definition \Eq{QEAG} for $W[\tilde{g},J]$, we find
\ba
\rme^{\rmi \Gamma[\tilde{g},\bar{g}] +  \rmi \int_{\tilde g} \bar{g}_{\mu\nu} J^{\mu\nu} } &= &
\int \cD h_{\mu\nu} \cD \bar{C}_\alpha \cD C^\beta \,  \cD b^\gamma 
\exp \Big(\rmi \Big\{S[\bar{g} + h]  \nn
&& +\frac{1}{2}  \int_{\tilde{g}}  t_\alpha^{\mu\nu}[\tilde{g}] (\bar{g}_{\mu\nu} + h_{\mu\nu} )
\, \cG^{\alpha \beta}[\tilde{g}] \, t_\beta^{\rho \sigma}[\tilde{g}]  ( \bar{g}_{\rho \sigma} + h_{\rho \sigma} )\nn
&& +\int_{\tilde{g}}  \bar{C}_\alpha \, t^{\alpha\mu\nu} [\tilde{g}]  \,r_{\mu\nu\beta} [\bar{g} + h] \,  C^\beta
+ \frac{1}{2}  \int_{\tilde{g}}  b_\alpha \cG^{\alpha \beta}[\tilde{g}] \, b_\beta\nn
& & +\int_{\tilde{g}} ( \bar{g}_{\mu\nu} + h_{\mu\nu}) \, J^{\mu\nu} \Big\}\Big).\label{ZJgh}
\ea
The second exponential on the left-hand side matches the next-to-last one on the right-hand side. Moreover, the last exponential on the right-hand side can be expressed in terms of the quantum effective action using \Eqq{extJ}:
\ba
\hspace{-0.8cm} 
\rme^{\rmi \Gamma[\tilde{g},\bar{g}] } &= &
\int \cD h_{\mu\nu} \cD \bar{C}_\alpha \cD C^\beta \cD b^\gamma  
\exp \Big(\rmi \Big\{ S[\bar{g} + h]- \!  \int_\eta  \frac{\delta \Gamma[\tilde{g},\bar{g}]}{\delta \bar{g}_{\mu\nu}} h_{\mu\nu} \nn
&&+  \frac{1}{2} \int_{\tilde{g}} \left[  t_\alpha^{\mu\nu}[\tilde{g}] \, (\bar{g}_{\mu\nu} + h_{\mu\nu} ) \right] \, \cG^{\alpha \beta}[\tilde{g}] \, \left[ t_\beta^{\rho \sigma}[\tilde{g}] \,  ( \bar{g}_{\rho \sigma} + h_{\rho \sigma} ) \right]\nn
&&+ \int_{\tilde{g}}  \bar{C}_\alpha \, t^{\alpha \mu\nu} [\tilde{g}]  \,r_{\mu\nu\beta} [\bar{g} + h] \,  C^\beta
+ \frac{1}{2} \int_{\tilde{g}}  b_\alpha \cG^{\alpha \beta}[\tilde{g}] \, b_\beta\,\Big\}\Big) ,
\label{quasiFinal}
\ea
where the first spacetime integral is $\int_\eta\coloneqq\int \rmd^D x$ because the $1/\sqrt{|\tilde{g}|}$ prefactor in \Eqq{extJ} cancels the one in the integral measure.

Finally, we identify $\tilde{g}_{\mu\nu}=\bar{g}_{\mu\nu} $ and we denote 
\be
\Gamma[\bar{g}]\coloneqq \Gamma[\tilde{g},\bar{g}]\Big|_{\tilde{g}=\bar{g}}\,.\label{finalQA}
\ee
The functional $\Gamma[\bar{g}]$ is the quantum effective action computed in a special gauge depending on the mean field $\bar{g}$. 
From \Eqq{quasiFinal} and the definition (\ref{finalQA}), it follows that
	\ba
	\rme^{\rmi\Gamma[\bar{g}] } &= &
	\int \cD h_{\mu\nu} \cD \bar{C}_\alpha \cD C^\beta \cD b^\gamma 
	\exp \Big(\rmi \Big\{ S[\bar{g} + h]-  \int_\eta
	\frac{\delta \Gamma[\tilde{g},\bar{g}]}{\delta \bar{g}_{\mu\nu}}\Big|_{\bar{g}= \tilde{g}}h_{\mu\nu}    \nn
	&&+  \frac{1}{2}  \int_{\bar{g}}  \left[ t_\alpha^{\mu\nu}[\bar{g}] \, (\bar{g}_{\mu\nu} + h_{\mu\nu} ) \right]
	\, \cG^{\alpha \beta}[\bar{g}] \, \left[ t_\beta^{\rho \sigma}[\bar{g}] \,  ( \bar{g}_{\rho \sigma} + h_{\rho \sigma} ) \right]\nn
	&& + \! \int_{\bar{g}} \! \bar{C}_\alpha \, t^{\alpha\mu\nu} [\bar{g}]  \,r_{\mu\nu\beta} [\bar{g} + h] \,  C^\beta
	+ \frac{1}{2}  \int_{\bar{g}}  b_\alpha \cG^{\alpha \beta}[\bar{g}] \, b_\beta\Big\}\Big).\label{FinalZ}
	\ea

We can slightly simplify the computations imposing a derivative gauge condition, namely, $t_\alpha^{\mu\nu}[\tilde{g}]$ in \Eqq{xi1} is a derivative operator with respect to the background metric $\tilde{g}_{\alpha \beta}$ and acting on the metric $g_{\alpha \beta}$. Therefore, the gauge functional in \Eqq{xi1} simplifies to 
\be
\chi_\alpha[\tilde{g},h] = t_\alpha^{\mu\nu} [\tilde{g}] \, h_{\mu\nu} \, ,
\label{xi2}
\ee
as a consequence of $\tilde{\nabla}_{\alpha} \tilde{g}_{\beta \gamma} =0$. Notice that the gauge-fixing condition is linear in the field $h_{\mu\nu}$. The ghost operator $M_\alpha^\beta[\tilde{g},g]$ \Eq{Mfirst} explicitly depends on $\tilde{g}$ and $h$, 
\be
M^\alpha{}_\beta[\tilde{g},h] = \frac{\delta \chi^\alpha}{\delta g_{\mu\nu}} \frac{\delta g_{\mu\nu}}{\delta \zeta^\beta}= t^{\a\mu\nu} [\tilde{g}]  \,r_{\mu\nu\beta} [\tilde{g},h] \, , 
%
\ee
where we replaced $\nabla$ in \Eqq{infDiff} with $\tilde{\nabla}$ because the generator $r_{\mu\nu\alpha}$ is at most linear in $g_{\mu\nu}$:
\ba
\delta g_{\mu \nu} &=&  r_{\mu\nu\alpha} [\tilde{g},h] \, \delta \zeta^\alpha \, , \nn
r_{\mu\nu\alpha}[\tilde{g}, h] &=& 
g_{\mu \alpha} \tilde{\nabla}_\nu + g_{\nu \alpha} \tilde{\nabla}_\mu 
= \tilde{g}_{\mu \alpha} \tilde{\nabla}_\nu +  h_{\mu \alpha} \tilde{\nabla}_\nu
+ \tilde{g}_{\nu \alpha} \tilde{\nabla}_\mu + h_{\nu \alpha} \tilde{\nabla}_\mu \,.
\ea
Using the gauge-fixing functional \cite{Asorey:1996hz} (in our signature)
\be
t_\alpha^{\mu\nu} [ \tilde{g} ]  
= \tilde{g}^{\mu \sigma}  \delta^\nu_\alpha \, \tilde{\nabla}_\sigma - \la \,  \tilde{g}^{\mu\nu} \tilde{\nabla}_\alpha\,,
\ee
where $\b_g$ is a constant, the gauge function reads
\ba
\chi_\alpha[\tilde{g},h]  = t_\alpha^{\mu\nu} [\tilde{g}] \, h_{\mu\nu} = \left( \tilde{g}^{\mu \sigma}  \delta^\nu_\alpha \, \tilde{\nabla}_\sigma - \la \,  \tilde{g}^{\mu\nu} \tilde{\nabla}_\alpha \right) h_{\mu\nu} 
 = {\tilde{\nabla}}_{\sigma} h^{\sigma}_{\alpha} - \la  {\tilde{\nabla}}_{\alpha} h \,.
\ea
Similarly, we can derive the ghost operator 
\ba
M^\alpha{}_\beta[\tilde{g},h] 
&= &   t^{\alpha\mu\nu} [\tilde{g}]  \,r_{\mu\nu\beta} [\tilde{g},h] \nn
& = & { \tilde \Box \delta^\alpha_\beta+ \tilde{\nabla}_{\beta} \tilde{\nabla}^{\alpha} 
	- 2  \la  \tilde{\nabla}^{\alpha} \tilde{\nabla}_{\beta}}   + 
\tilde{\nabla}^\mu \left( h_{\mu \beta} \tilde{\nabla}^\alpha \right) \nn
&&+ \tilde{\nabla}^\mu \left( h^\alpha_{\beta} \tilde{\nabla}_\mu  \right) 
- 2 \la \tilde{\nabla}^\alpha \left( h_{\mu\beta} \tilde{\nabla}^\mu  \right), \label{malf}
\ea
which consists in a kinetic operator (third line of \Eqq{malf}) and few a interaction terms all linear in the graviton. Note that the ghost nature of the field $C^\a$ is not due to a wrong sign in front of the kinetic term (the kinetic matrix $M^\alpha{}_\beta$ has the canonical sign for a complex scalar) but to the fact that $C^\a$ is a Grassmann variable. Overall,\index{gauge fixing!action}
\ba
S_{\rm gf} & = &
	\frac{1}{2} \int \! \rmd^D x \sqrt{|\tilde{g}|} \, \chi_{\alpha}[\tilde{g},h] \,  \cG^{\alpha \beta}[\tilde{g}] \, \chi_{\beta}[\tilde{g},h] \, ,  
	\label{GFINfunc} \\ 
	&&
	\chi_{\alpha} [\tilde{g},h] = {\tilde{\nabla}}_{\sigma} h^{\sigma}_{\alpha} - \la  {\tilde{\nabla}}_{\alpha} h \,,\label{chia}\\
	&& 
	\cG^{\mu \nu}[\tilde{g}] = - \frac{\Mpl^2}{4} \left[2\tilde{g}^{\mu \nu}(\la_1+\g_2'\la_3\tilde\B)
	+\g_2'\la_2 \tilde\N^\mu\tilde\N^\nu -2\g_2'\la_3\tilde\N^\nu\tilde\N^\mu\right],\label{shapiro3ain} \\
	\hspace{-1cm} 
	S_{\rm gh} & = & \int \! \rmd^D x \sqrt{|\tilde{g}|} \left[  \bar{C}_{\alpha} \, M^\alpha{}_\beta[\tilde{g},h] \, C^{\beta} 
	+\frac{1}{2}  b_{\alpha} \cG^{\alpha \beta}[\tilde{g}] b_{\beta} \right] \, , 
	\label{tuttighosts} \\
	&&
	M^\alpha{}_\beta[\tilde{g},h] := M^\alpha{}_\beta[\tilde{g}] + M^\alpha{}_\beta[\bar{C}, C, h]\,,  \\
	&& 
	M^{\alpha}{}_{\beta}[\tilde{g}] = \tilde{\Box} \delta^{\alpha}_{\beta}  
	+{\tilde{\nabla}}_{\beta} {\tilde{\nabla}}^{\alpha}  - 2 \la {\tilde{\nabla}}^{\alpha} {\tilde{\nabla}}_{\beta}\,,\label{Mab}\\
	&&
	M^\alpha{}_\beta[\bar{C}, C, h] = \tilde{\nabla}^\mu \left( h_{\mu \beta} \tilde{\nabla}^\alpha \right) 
	+ \tilde{\nabla}^\mu \left( h^\alpha_{\beta} \tilde{\nabla}_\mu  \right) 
	- 2 \la \tilde{\nabla}^\alpha \left( h_{\mu\beta} \tilde{\nabla}^\mu  \right), \label{shapiro3a}
	\ea
where $\g_2'=\g_2'(\tilde\B)$ and the gauge-fixing parameters\index{gauge fixing!parameters} $\la$, $\la_1$, $\la_2$ and $\la_3$ are dimensionless constants, $[\la]=[\la_i]=0$, just like for a local theory.\footnote{As we will see in section \ref{secpi}, there is no need to generalize $\la$ and $\la_i$ to non-trivial operators.} In \Eqq{GFINfunc}, we used the compatibility condition $\tilde{\nabla}_\alpha \, \tilde{g}_{\beta \gamma}= 0$ for the background metric. Note that, on a curved background, covariant derivatives do not commute, hence the separation of the last two factors in \Eq{shapiro3ain}. The expression for $\cG^{\mu \nu}$ generalizes the one of  \cite{Asorey:1996hz} to the case of nonlocal theories.

The action for the Faddeev--Popov ghosts\index{Faddeev--Popov ghosts} only has two derivatives (\Eqq{Mab}) but it can be modified in order to have the same number of derivatives as in the gravitational and in the gauge-fixing action in the UV, which is higher-order but finite in nonlocal quantum gravity with asymptotically polynomial operators. This can be done introducing the identity in the path integral in the form of $(\det \, \cG^{\alpha \beta})^{-1/2} \times (\det \, \cG^{\alpha \beta})^{1/2}$ \cite{Fradkin:1981iu,Asorey:1996hz}.

To summarize, the quantum effective action is given by the path integral \Eq{FinalZ} with the gauge-fixing action \Eq{GFINfunc}, the ghost action \Eq{tuttighosts} and with the metric $\tilde{g}_{\alpha \beta}$ identified with the background metric $\bar{g}_{\alpha \beta}$. 


\subsection{Loop expansion}

In the previous subsection, we derived the path-integral formula \Eq{FinalZ} for the quantum effective action. However, in order to solve such equation for $\Gamma$ we recall a perturbative technique that goes under the name of loop expansion. 

We expand the functional $S[\bar{g} + h]$ as a Taylor series of the field $\bar{g}_{\mu\nu}$,
\bs\ba
S[\bar{g} + h]   &=& S[\bar{g}] + \sum_{n=1}^{\infty} \frac{1}{n!}S_n[\bar{g}] h^n\,,\\
S_n[\bar{g}] h^n &\coloneqq& \int_{\eta_1} \ldots \int_{\eta_n}\,\frac{\de S[\bar{g}]}{\de\bar{g}_{\mu\nu}(x_1)\ldots\de\bar{g}_{\s\t}(x_n)}\,h_{\mu\nu}(x_1)\ldots h_{\s\t}(x_n)\,,
\ea\es
where $\int_{\eta_n}\coloneqq\int\rmd^D x_n$, and define
\be
\Gamma_1[\bar{g}] h \coloneqq \int_\eta\, \left.\frac{\delta \Gamma[\tilde{g},\bar{g}]}{\de\bar{g}_{\mu\nu}(x)}\right|_{\bar{g}= \tilde{g}}\,h_{\mu\nu}(x)\,.
\ee
Equation \Eq{FinalZ} can be recast as
\ba
\rme^{\rmi\bar\Gamma[\bar{g}]} &= &
\int \cD h_{\mu\nu} \cD \bar{C}_\alpha \cD C^\beta \cD b^\gamma 
\exp \Big(\rmi \Big\{\frac{1}{2}S_2[\bar{g}] h^2+\sum_{n=3}^{\infty} \frac{1}{n!}S_n[\bar{g}] h^n +(S_1[\bar{g}]-\Gamma_1[\bar{g}]) h   \nn
&&+  \frac{1}{2}  \int_{\bar{g}}  \left[ t_\alpha^{\mu\nu}[\bar{g}] \, (\bar{g}_{\mu\nu} + h_{\mu\nu} ) \right]
\, \cG^{\alpha \beta}[\bar{g}] \, \left[ t_\beta^{\rho \sigma}[\bar{g}] \,  ( \bar{g}_{\rho \sigma} + h_{\rho \sigma} ) \right] 
\nn
&& + \! \int_{\bar{g}} \! \bar{C}_\alpha \, t^{\alpha\mu\nu} [\bar{g}]  \,r_{\mu\nu\beta} [\bar{g} + h] \,  C^\beta
+ \frac{1}{2}  \int_{\bar{g}}  b_\alpha \cG^{\alpha \beta}[\bar{g}] \, b_\beta\Big\}\Big).\label{FinalZ2}
\ea
where $\bar\Gamma[\bar{g}]\coloneqq \Gamma[\bar{g}]-S[\bar{g}]$ encodes all the quantum corrections augmenting the classical action $S[\bar{g}]$. In the right-hand side, we separated explicitly the quadratic term defining the propagator from non-linear interactions. The term $S_1-\Gamma_1$ cancels one-particle-reducible diagrams coming from other contributions in the expression, so that the right-hand side only contains one-particle-irreducible diagrams \cite{BOS}.


\subsection{One-loop effective action} \label{1LoopQG}

In order to derive the one-loop quantum effective action, we have to expand the action at the second order in the  fields. In particular, we expand the gravitational action at the second order in $h_{\mu\nu}$, so that no gravitational coupling appears between the ghosts and the graviton. Therefore, \Eq{FinalZ} simplifies to 
\ba
\rme^{ \rmi \Gamma[\bar{g}] } &= &
\int \cD h_{\mu\nu} \cD \bar{C}_\alpha \cD C^\beta \cD b^\gamma 
\exp \rmi \Bigg[ S[\bar{g}] + \frac12 h_{\mu\nu} \,\,  \frac{\delta^2{S[\bar{g} + h]}}{{\delta h_{\mu\nu} \, \delta h_{\rho \sigma}} }\Bigg|_{h=0} \!h_{\rho \sigma} \nn
&& +\frac{1}{2} \int_{\bar{g}} \chi_{\alpha}[\bar{g},h] \, \cG^{\alpha \beta}[\bar{g}] \, \chi_{\beta}[\bar{g},h] + \int_{\bar{g}} \left( \bar{C}_{\alpha} \, M^\alpha{}_\beta[\bar{g}] \, C^{\beta} 
+\frac{1}{2}  b_{\alpha} \cG^{\alpha \beta}[\bar{g}] b_{\beta} \right)\Bigg]\nn
& = & \rme^{\rmi S[\bar{g}]}  \, \left( \det \, \Hes^{\mu\nu \rho\sigma} \right)^{-\frac{1}{2}}
(\det  \, M^{\alpha}{}_{\beta} ) \, \, 
(\det  \, \cG^{\mu \nu} )^{\frac{1}{2}}  , \label{Gamma1Loop}
\ea
where the gauge-fixing and the ghost operators are given by \Eqqs{chia} and \Eq{Mab}, respectively, while the Hessian is\index{Hessian}
\be
\Hes^{\mu\nu \rho \sigma} \coloneqq 
\frac{\delta^2 S[\bar{g} + h]}{ \delta h_{\mu \nu}\delta h_{\rho \sigma}}\Bigg|_{h=0} 
\!\!\! 
+  \frac{\delta \chi_{\alpha}}{\delta h_{\mu \nu}} \, \cG^{\alpha\beta}[\bar{g}]  \, 
\frac{\delta \chi_{\beta}}{\delta h_{\rho \sigma}}\Bigg|_{h=0} ,
\label{H}
\ee
where $\cG^{\alpha \beta}[\bar{g}]$ is given in \Eqq{shapiro3ain}. 

Taking $- \rmi \ln$ of \Eqq{Gamma1Loop}, we finally get the quantum effective action at one loop,
\ba
\Gamma^{(1)}[g]  =  - \rmi  \ln  \rme^{ \rmi \Gamma[\bar{g}] }  =S[g] + \frac{\rmi}{2}   \ln \det \cH - \rmi  \ln \det {M}
- \frac{\rmi}{2} \ln\det {C}  . \label{1loopGamma}
\ea

The effective action in nonlocal quantum gravity has not been calculated in full yet and super-renormalizability has been checked so far with the power-counting argument and via the calculation of the one-loop effective action in scalar toy models and in gravitational models with fewer derivatives than those appearing in the UV limit of asymptotically polynomial form factors, in particular, six \cite{Rachwal:2021bgb}. However, increasing the number of derivatives makes the power counting even more powerful because the beta functions do not depend on the running couplings. In general, divergences come only from the polynomial (local) parts of the theory, while diagrams with nonlocal form factors within are, by definition, all convergent on the domain of such operators. It is important to stress that power counting is valid, as a general argument, to conclude that the theory is super-renormalizable. The reason is that the counter-terms to be added to the bare Lagrangian are local operators \cite{Calcagni:2023goc}, which implies that the Bogoliubov--Parasiuk--Hepp--Zimmermann (BPHZ) subtraction scheme holds and that the theory is also BPHZ renormalizable \cite{Calcagni:2022shb}. This conclusion is somewhat obvious from the fact that the UV limit of nonlocal theories with asymptotically polynomial form factors is local but, surprisingly, it also holds for some nonlocal theories with a nonlocal UV limit such as fractional gravity \cite{Calcagni:2022shb}. An immediate consequence is that BPHZ renormalization guarantees that sub-divergences can always be absorbed by a standard one-loop counter-term in the action. Hence, the power-counting argument is sufficient to account for all divergences at any loop order. Note also that these complications do not appear in the finite version of the theory. In this case, there are no sub-divergences at higher loop orders because there are no divergences at all at lower order, since all beta functions vanish \cite{Modesto:2014lga,Calcagni:2023goc}. Thanks to these general results, the explicit calculation of the one-loop quantum effective action \Eq{1loopGamma} becomes secondary in the discussion of the renormalizability of nonlocal quantum gravity, although, of course, it should be on top of the future agenda.


\subsection{Green's functions}\label{grefu}

This subsection is devoted to the path-integral formulation of any gravitational theory in Minkowski spacetime. Namely, the background spacetime is globally Minkowski. Here we treat gravity in exactly the same way as the other fundamental interactions in Nature. Indeed, the choice of the Minkowski background in gravity is analogue to the trivial zero background for gauge bosons, scalars, or fermions and it is dictated by the equivalence principle.

The Green's functions for quantum gravity are related to the generating functional $Z[J]$ through 
multiple derivatives with respect to the source $J_{\mu\nu}$:
\ba
\hspace{-1cm}
G_{\mu_1 \nu_1 \mu_2 \nu_2 \dots \mu_n \nu_n}(x_1, \dots, x_n ) &\coloneqq & 
\langle h_{\mu_1 \nu_1}(x_1)   \dots h_{\mu_n \nu_n}(x_n ) \rangle \nn
&=& \frac{1}{\rmi^n} \frac{\delta^n Z[J]}{\delta J^{\mu_1 \nu_1}(x_1) \, \delta J^{\mu_2 \nu_2}(x_2) \dots J^{\mu_n \nu_n}(x_n) }\Bigg|_{J=0},\label{Ghz}
\ea
where the right-hand side has to be evaluated in $J=0$. The generating functional is obtained simply replacing $\tilde{g}_{\mu \nu} = \eta_{\mu\nu}$ in \Eqq{Zgtilde}: 
\ba
\hspace{-1cm}
Z[\eta,J] \! &=& \rme^{\rmi W[\eta,J]}  \nn
&=&  \int  \cD h_{\mu\nu} \, \cD \bar{C}_\alpha {\rm} \cD C^\beta \cD b^\gamma \, 
\rme^{\rmi \left(S[g] + S_{\rm gf}[\eta,g] + S_{\rm gh}[\eta,g,C,\bar{C},b] + 
	\int_\eta h_{\mu\nu} J^{\mu\nu}\right) } , 
\label{ZgtildeMink} 
\ea 
where $g_{\mu\nu} = \eta_{\mu\nu} + h_{\mu\nu}$, while the gauge-fixing and ghost operators (\ref{gaugefixing}) and (\ref{ghostsCb}) simplify to\index{gauge fixing!action}\index{ghost!action}
\ba
&& \hspace{-0.5cm}
S_{\rm gf}[\eta,h]  = \frac{1}{2}
\int_\eta  \chi_\alpha[h, \eta] \, \cG^{\alpha \beta}[\eta] \, \chi_\beta[h, \eta]
\, , \label{gaugefixingh} \\
&& \hspace{-0.5cm}
S_{\rm gh}[h,C,\bar{C},b] = \underbrace{\int_\eta \bar{C}_\alpha M^\alpha_\beta[\eta,h] C^\beta}_{S_{\rm gh}[\eta,h,C,\bar{C}]}
+\underbrace{ \frac{1}{2} \int_\eta b_\alpha\cG^{\alpha \beta}[\eta] b_\beta}_{S_{\rm gh}[\eta,b]}\label{ghostsCbh}\,.
\ea

In perturbation theory, we split the action into a kinetic and an interacting part. The kinetic Lagrangian is quadratic in the perturbation $h_{\mu\nu}$, while interactions\index{interactions} are at least cubic in $h_{\mu\nu}$. Let us rename the action for the free theory $S_0[h]$ and reserve the subscript 0 for any quantity evaluated with the free dynamics. Therefore, the generating functional for the free theory reads
\ba
\hspace{-0.8cm}
Z_0[J]
\label{Zgtilde0}
&=&  \int  \cD h_{\mu\nu} \cD \bar{C}_\alpha \cD C^\beta \cD b^\gamma \, 
\rme^{\rmi \left(S_0[h] + S_{\rm gf}[h] + S_{\rm gh}[C,\bar{C},b] + 
	\int_\eta h_{\mu\nu} J^{\mu\nu}\right) }, 
\ea
where we have omitted the background Minkowski metric from the arguments of the functionals.

Since ghosts cannot appear as asymptotic states, we did not include source terms for such fields in \Eqq{Zgtilde}. However, in formulating perturbation theory it is convenient to consider also Green's functions involving ghost fields. The generating functional \Eq{Zgtilde} is extended to
\ba
Z[J, J_C, J_{\bar{C}}, J_b] &=&
\int  \cD h_{\mu\nu} \cD \bar{C}_\alpha \cD C^\beta \cD b^\gamma \,\exp \rmi \Big[  S[g] + S_{\rm gf}[h] + S_{\rm gh}[h,C,\bar{C},b]\nn
&& \qquad + \int_\eta \left(  h_{\mu\nu} J^{\mu\nu} + J_{C, \alpha} C^\alpha+ J^{\bar{C},\alpha} \bar{C}_\alpha
+ J^{b, \alpha} b_\alpha \right)\Big],\label{Zghosts}
\ea 
where the gauge-fixing action\index{gauge fixing!action} and the actions for the ghosts\index{ghost!action} are obtained from \Eqqs{GFINfunc}, (\ref{shapiro3ain}), (\ref{tuttighosts}) and (\ref{shapiro3a}) after replacing $\tilde{g}_{\alpha \beta}$ with the Minkowski metric $\eta_{\alpha \beta}$.

Therefore, the Green's functions are
\ba
&& \hspace{-0.6cm}\langle h_{\mu \nu}(x_1)   \dots   \bar{C}_\alpha (y_1)     \dots  {C}^\beta (z_1)    \dots b^\gamma (w_1) \dots \rangle  \nn
&& = \frac{1}{\rmi^n} \frac{\delta^n Z[J, J_C, J_{\bar{C}}, J_b]}{\delta J^{\mu \nu}(x_1) \dots \delta J^{\bar{C}, \alpha }(y_1) \dots \delta J_{{C}, \beta} (z_1) \dots \delta J^{b, \gamma} (w_1) \dots}\Bigg|_{J=0},
\label{GwGhosts}
\ea
where the notation $\langle \dots \rangle$ means that we are integrating \Eqq{Zghosts} with insertions of graviton and ghosts fields,
\ba
&& \hspace{-.6cm}\langle h_{\mu \nu}(x_1)   \dots   \bar{C}_\alpha (y_1)     \dots  {C}^\beta (z_1)    \dots b^\gamma (w_1) \dots \rangle  \nn
&& =  \int  \cD h_{\mu\nu} \cD \bar{C}_\alpha \cD C^\beta \cD b^\gamma \, h_{\mu \nu}(x_1)   \dots   \bar{C}_\alpha (y_1)     \dots  {C}^\beta (z_1)    \dots b^\gamma (w_1) \dots \nn
&& \hspace{1.2cm}    \times \exp \rmi \Big[  S[g] + S_{\rm gf}[h] + S_{\rm gh}[h,C,\bar{C},b]  \nn
&& \hspace{1.2cm} +  \int_\eta \left(  h_{\mu\nu} J^{\mu\nu} + J_{C, \alpha} C^\alpha+ J^{\bar{C},\alpha} \bar{C}_\alpha
+ J^{b, \alpha} b_\alpha \right)\Big].\label{insertions}
\ea
It is straightforward to prove that \Eqq{insertions} is simply the multiple functional derivative of the generating  functional $Z$ with  respect to the currents $\rmi J$, as stated by \Eqq{GwGhosts}.
 

\subsection{Efimov analytic continuation}\label{efi}

It has become progressively clear that even the best behaved nonlocal quantum field theories have problems if scattering amplitudes are calculated directly in Lorentzian signature \cite{Briscese:2021mob}. In other words, usually, naive Lorentzian nonlocal theories where the internal energies $k^0$ are integrated on the real line do not exist. If, instead, one defines the Feynman diagrams with imaginary external and internal energies (Euclidean signature in momentum space) and afterwards analytically continues external energies to real ones after integrating, then loop integrals can be performed consistently and the theory admits a unique Lorentzian limit. This is the so-called Efimov analytic continuation \cite{Efimov:1967dpd,Pius:2016jsl,Briscese:2018oyx,Chin:2018puw,Koshelev:2021orf,Buoninfante:2022krn}, which consists of three steps. 

For any given loop Feynman diagram in Lorentzian quantum gravity, namely, where the path integral is performed on Lorentzian metrics (perturbations on Minkowski spacetime) with the $\rmi\e$ Feynman prescription:
\begin{enumerate}
\item Assume that both internal and external energies (respectively, $k^0$ and $p^0$) take complex values in the loop amplitudes.	
\item Integrate on the imaginary axis in the $(\Re\,k^0,\,\Im\,k^0)$ complex plane. (Note that, in general, there will be poles $\bar k^0(p)$ of the integrand to the left and to the right of the path, but none on the path itself \cite{Briscese:2018oyx}.)
\item After the loop integrations, analytically continue the external energies $p^0$ back to real values.
\end{enumerate}
We can express the above three steps into other, equivalent ways. 

One, which we might call \emph{deformed-paths view}, is to assume that only the internal energies $k^0$ are complex valued, while the external energies $p^0$ stay real. Then, internal energies are integrated along special paths in the complex plane. According to the previous description, these paths are obtained moving analytically the external energies from complex to real values. Each path is deformed in such a way as to accommodate analytically the poles that migrate across the imaginary $k^0$ axis, thus guaranteeing safe integration \cite{Briscese:2018oyx}. The advantage of understanding Efimov analytic continuation according to steps 1-3 instead of in the deformed-paths view is that the correct deformed path and the homotopic class it represents can hardly be guessed if external energies stay real and internal ones are imaginary. 

Another equivalent form (\emph{Euclidean theory view}) is to invoke the Euclidean version of the theory, which corresponds to the third line of \Eq{scheme} below. In step 2, define first a Euclidean version $k_D = -\rmi k^0$ and $p_D = -\rmi p^0$ of both internal and external energies and integrate on real values of $k_D$. Obviously, in Euclidean quantum gravity this step would be automatically implemented. Then, it is clear that the Lorentzian amplitudes of the nonlocal theory are nothing but Euclidean amplitudes analytically continued to imaginary external energies $p_D$. Therefore, \emph{at the perturbative level}, the Lorentzian theory as defined in sections \ref{qeasec}--\ref{grefu} is indistinguishable from an analytically continued Euclidean version of the theory. 

Of course, this means that the Lorentzian and the pure Euclidean theory (i.e., without analytic continuation) are physically inequivalent, as also noted in \cite{Buoninfante:2022krn}, since there is a highly non-trivial analytic continuation differentiating the two. In the pure Euclidean case, one integrates straight on top of the real axis in the $(\Re\,k_D,\Im\,k_D)$ plane, and this is it. In contrast, in the Lorentzian case one deforms this path when sending the external energies to real values and some poles get across the imaginary axis \cite{Briscese:2018oyx}. This is also the reason why we started with a Lorentzian path-integral formulation instead of a Euclidean one: the final result is unique because the power counting and all the integrals are defined in Euclidean momentum space.

Summarizing, at the perturbative level there is no conceptual difference between gravity and any other interactive gauge theory. Perturbative quantum gravity is the theory of interacting gravitons and loop amplitudes can be computed exactly as in those cases. All loop amplitudes are computed with purely imaginary internal and external energies and, in the end, one makes the above analytic continuation and the final result is a Lorentzian quantum effective action. The only difference with respect to two-derivative theories is that the analytically continued theory is not equivalent to the theory defined with internal real energies because of the contribution of the pole at infinity (essential singularity) due to the nonlocal form factor.


\section{Tree-level graviton propagator}\label{sec3}

In this subsection, we calculate the tree-level graviton propagator for the general action \Eq{GA} with arbitrary form factors $\g_{0,2,4}$, following the same steps as in \cite{Accioly:2002tz}. This will be central to the perturbative solution of the conformal instability problem.

The tree-level propagator for the dimensionless gravitational perturbation $h_{\mu\nu}$ defined by
\be\label{gh}
g_{\mu\nu}=\eta_{\mu\nu}+h_{\mu\nu}\,,
\ee
can be computed in a standard albeit lengthy way. The expansion at second order of the action \Eq{GA} yields the Lagrangian\footnote{Here is how to compare the conventions and results of  \cite{Accioly:2002tz} with ours. Spacetime signature is mostly minus in  \cite{Accioly:2002tz} and mostly plus for us. Also, their Ricci scalar $R$ is $-R$ for us, so that there is an extra overall $-$ sign in the total action and in its coefficients $\a=-\Mpl^2\g_0$, $\b=-\Mpl^2\g_2$ and $\g=-\Mpl^2\g_4$. Then, $b=\b\k^2/2=-2\g_2$, $c=\a/\b=\g_0/\g_2$ and $d=\g\k^2/2=-2\g_4$, where $\k=2/\Mpl$. Finally, the graviton in  \cite{Accioly:2002tz} is dimensionful and defined as $g_{\mu\nu}=\eta_{\mu\nu}+\k h_{\mu\nu}$, while our $h_{\mu\nu}$ is dimensionless. Therefore, our coefficients $c_i\in \{c_1,c_2,c_0,\bar c_0,\bar{\bar{c}}_0\}$ are related to the coefficients $x_i\in \{x_1,x_2,x_0,\bar x_0,\bar{\bar{x}}_0\}$ in  \cite{Accioly:2002tz} by $c_i(k^2)=-(\Mpl^2/4)\, x_i(k^2)$. The expressions in \cite{Accioly:2002tz} immediately generalize to the weakly nonlocal case with non-trivial form factors and we have recalculated them anew.}
\ba
\cL_{\rm K} &=& \frac{\Mpl^2}{8} \Big\{h^{\mu \nu} \B h_{\mu \nu} + A_{\mu}A^\mu+(A_{\mu} - \p_\mu\vp)(A^{\mu}- \p^\mu\vp)  \nn
&& \quad\qquad +\B h_{\mu \nu}  \g_2'(\B) \B h^{\mu \nu}-(\p_\mu A^{\mu}) \g_2'(\B) \p_\nu A^{\nu}- F^{\mu \nu}\g_2'(\B) F_{\mu \nu} \nn
&&\quad\qquad +(\p_\mu A^{\mu}-\B\vp) [\g_2'(\B)+ 4\g_0'(\B)] (\p_\nu A^{\nu}-\B\vp)\Big\}\nn
&\eqqcolon& \frac{1}{2} \,h^{\a \b}  \,  \cO^{\rm K}_{\a\b\mu\nu}\, h^{\mu \nu}+\dots\,,\label{quadratic2}
\ea
where $A^{\mu} \coloneqq \p_\nu h^{\mu \nu}$, $F_{\mu \nu}\coloneqq \p_\mu A_\nu-\p_\nu A_\mu$,  $\vp \coloneqq h^{\mu}_{\ \mu}$, $\g_{0,2}'$ are defined in \Eq{gprime} and dots denote total derivative terms. Since the kinetic term $ \cO^{\rm K}$ is not invertible, one must add to $\cL_{\rm K}$ the gauge-fixing Lagrangian in \Eq{GFINfunc}. On Minkowski background $\tilde g^{\mu\nu}=\eta^{\mu\nu}$, the covariant derivatives in \Eq{shapiro3ain} become ordinary commuting derivatives. Since $\chi_\mu = A_\mu-\la\p_\mu \vp$, we obtain
\ba
\mathcal{L}_{\rm gf} &=& \frac{\Mpl^2}{8} \Big[
-2\la_1 (A_\mu - \la\p_\mu\vp) ( A^\mu - \la \p^\mu\vp)\nn
&&\quad\qquad+\la_2(\p_\a A^\a - \la \B \vp)\g_2'(\B)(\p_\b A^\b - \la\B \vp)+\la_3F_{\mu\nu} \g_2'(\B) F^{\mu\nu}\Big]\nn
&\eqqcolon& \frac{1}{2} \,h^{\a \b}  \,  \cO^{\rm gf}_{\a\b\mu\nu}\, h^{\mu \nu}+\dots\,.
\ea
In momentum space, $\B=-k^2=(k^0)^2-\bm{k}^2$. In Euclidean momentum space, $=-k^2=-k_D^2-\bm{k}^2\leq 0$.

Contrary to $\cO^{\rm K}$, the kinetic operator $\cO=\cO^{\rm K}+\cO^{\rm gf}$ is invertible. In the basis of Barnes--Rivers projectors \cite{Riv64,Bar65,VanNieuwenhuizen:1973fi} and in momentum space, it is
\be\label{E23}
{\cO}(k)=c_{1}{P}^{(1)}+c_{2}{P}^{(2)}+c_{0}{P}^{(0)}+\bar c_{0}\bar{{P}}^{(0)}+\bar{\bar{c}}_{0}\bar{\bar{{P}}}^{(0)},
\ee
where we omitted spacetime indices and
\bs\label{projectors}\ba
&& P^{(1)}_{\mu\nu\rho\s} \coloneqq \Theta_{\mu(\rho} \om_{\s)\nu} + \Theta_{\nu (\rho} \om_{\s)\mu}\,,\qquad \om_{\mu\nu} \coloneqq \frac{k_{\mu} k_{\nu}}{k^2}\,,\qquad \Theta_{\mu \nu} \coloneqq \eta_{\mu \nu} - \om_{\mu\nu}\,,\\
&& P^{(2)}_{\mu\nu\rho\s} \coloneqq \frac{1}{2} \left(\Theta_{\mu \rho} \Theta_{\nu \s} +  \Theta_{\mu\s} \Theta_{\nu\rho} \right)
- \frac{1}{D-1} \Theta_{\mu\nu} \Theta_{\rho\s} \, , \\
&&  P^{(0)}_{\mu\nu\rho\s} \coloneqq \frac{1}{D-1} \Theta_{\mu \nu} \, \Theta_{\rho \s}\,,\\
&&\bar{P}^{(0)} _{\mu\nu\rho\s} =  \om_{\mu \nu} \, \om_{\rho \s} \,,\label{proje22}\\
&&\bar{\bar P}_{\mu\nu\rho\s}^{(0)} =\left(\Theta_{\mu\nu}\om_{\rho\s}+\om_{\mu\nu}\Theta_{\rho\s}\right).
\ea\es
The coefficients $c_i$ are
\bs\label{E25}\ba
c_{1}  &=&  -\frac{\Mpl^{2}}{4} k^2\left(\la_1-k^2\la_3 \g_2'\right), \\
c_{2}  &=&  -\frac{\Mpl^{2}}{4} k^22\left(1-k^2\g_2'\right), \\
c_{0}  &=&  \frac{\Mpl^{2}}{4}k^2
\left\{(D-2)-2(D-1)\la^2\la_1+k^{2}\left[4(D-1)\g_0'+D\g_2'+(D-1)\la^2\la_2\g_2'\right] \right\}, \nn\\
\bar c_{0}  &=&  -\frac{\Mpl^{2}}{4}k^2(\la-1)^22\left(2\la_1-k^2\la_2\g_2'\right),\\
\bar{\bar{c}}_{0}  &=& -\frac{\Mpl^{2}}{4}k^2\la(\la-1)2\left(2\la_1-k^2\la_2\g_2'\right).
\ea\es
Note that $[\cO]=D=[c_i]$. In order to find the propagator
\be
{\cO}^{-1}(k)=s_{1}{P}^{(1)}+s_{2}{P}^{(2)}+s_{0}{P}^{(0)}+\bar s_{0}\bar{{P}}^{(0)}+\bar{\bar{s}}_{0}\bar{\bar{{P}}}^{(0)},
\ee
one has to solve the linear system 
\ben
{\cO} \cdot  {\cO}^{-1} = 
\begin{pmatrix}c_1 & 0 & 0 & 0 & 0\\
	0 & c_2 & 0 & 0 & 0\\
	0 & 0 & c_0 & 0 & \bar{\bar{c}}_{0}\\
	0 & 0 & \bar{\bar{c}}_{0} & 0 & \bar{c}_0\\
	0 & 0 & 0 & \bar{c}_0 & \bar{\bar{c}}_{0}\\
	0 & 0 & 0 & \bar{\bar{c}}_{0} & c_0
\end{pmatrix}\begin{pmatrix}s_{1}\\
	s_{2}\\
	s_{0}\\
	\bar{s}_0\\
	\bar{\bar{s}}_{0}
\end{pmatrix}=\begin{pmatrix}1\\
	1\\
	1\\
	0\\
	1\\
	0
\end{pmatrix}
.
\een
Using the echelon matrix form \cite{Accioly:2002tz}
\ben
\begin{pmatrix}c_{1} & 0 & 0 & 0 & 0 & 1\\
	0 & c_{2} & 0 & 0 & 0 & 1\\
	0 & 0 & c_0 & 0 & (D-1)\bar{\bar{c}}_{0} & 1\\
	0 & 0 & \bar{\bar{c}}_{0} & 0 & \bar{c}_0 & 0\\
	0 & 0 & 0 & \bar{c}_0 & (D-1)\bar{\bar{c}}_{0} & 1\\
	0 & 0 & 0 & \bar{\bar{c}}_{0} & c_0 & 0
\end{pmatrix}\stackrel{\rm R}{\sim}\begin{pmatrix}c_{1} & 0 & 0 & 0 & 0 & 1\\
	0 & c_{2} & 0 & 0 & 0 & 1\\
	0 & 0 & c_0 & 0 & (D-1)\bar{\bar{c}}_{0} & 1\\
	0 & 0 & 0 & \bar{c}_0 & (D-1)\bar{\bar{c}}_{0} & 1\\
	0 & 0 & 0 & 0 & c_0\bar{c}_0-(D-1)\bar{\bar{c}}_{0}^{2} & -\bar{\bar{c}}_{0}\\
	0 & 0 & 0 & 0 & 0 & 0
\end{pmatrix} \, ,
\een
where $\stackrel{\rm R}{\sim}$ denotes row equivalence, the graviton propagator is
\be\label{E27}
{\cO}^{-1}=\frac{1}{c_{1}}{P}^{(1)}+\frac{1}{c_{2}}{P}^{(2)}+\frac{1}{c_0\bar{c}_0-(D-1)\bar{\bar{c}}_{0}^{2}}\left[\bar{c}_0{P}^{(0)}+c_0\bar{{P}}^{(0)}-\bar{\bar{c}}_{0}\bar{\bar{{P}}}^{(0)}\right]\,.
\ee
All these expressions hold also in the case where $\g_0'$ and $\g_2'$ are nontrivial functions of the momentum as in the theory \Eq{GA}. Note that the coefficients in front of $P^{(2)}$ and $P^{(0)}$ do not depend on the gauge parameters and we can isolate a gauge-independent part ${\cO}_{\rm K}^{-1}$ in the propagator:
\ba
{\cO}^{-1}(k) &=& {\cO}_{\rm K}^{-1}(k) + \textrm{gauge}\nn
&=&\frac{1}{c_{2}}{P}^{(2)}+\frac{\bar{c}_0}{c_0\bar{c}_0-(D-1)\bar{\bar{c}}_{0}^{2}}{P}^{(0)} + \textrm{gauge}\nn
&=&- \frac{4}{\Mpl^{2}} \Bigg(\frac{{P}^{(2)}}{k^2 \left[1-k^2 \g_2' (k^2) \right]}-\frac{{P}^{(0)}}{k^2 \left\{(D-2) + k^2 \left[4(D-1) \g_0' (k^2) + D \g_2' (k^2)\right]\right\}}\Bigg) \nn
&&+ \textrm{gauge}\,.\label{GP}
\ea
This expression is equivalent to ${\cO}_{\rm K}^{-1}={P}^{(2)}/c_2+{P}^{(0)}/c_0$ when setting $\la_1=\la_2=\la_3=0$. In general, different gauge choices become handy when exploring different aspects of the theory, for instance, when showing that the spin-0 mode with the wrong sign does not propagate or when checking positivity of the eigenvalues of the kinetic operator in Euclidean signature. Some common gauge choices are the Julve--Tonin gauge ($\la=0$) \cite{Julve:1978xn}, the de Donder gauge ($\la=1/2$, $\la_2=0=\la_3$) and the Feynman gauge ($\la=1/2$, $\la_1=1$, $\la_2=0=\la_3$) \cite{Accioly:2002tz}.

The gauge-independent part of the propagator \Eq{GP} has already been studied in the literature. As its equivalent form shows, both in the minimally and in the nonminimally coupled version of nonlocal quantum gravity one has
\be
{\cO}^{-1}(k) =- \frac{4}{\Mpl^{2}} \Bigg[\frac{{P}^{(2)}}{k^2 \rme^{\H_2(k^2)}}-\frac{{P}^{(0)}}{(D-2)k^2\rme^{\H_0(k^2)}}\Bigg]+ \textrm{gauge}\,,\label{GP2}
\ee
which does not have extra poles beside from the usual ones. This conservation of the unitarity property is the physical motivation for choosing the form factors \Eq{gamma} and one of the main pillars of nonlocal quantum gravity.


\section{Conformal instability}\label{secpi}

Consider a generic theory of quantum gravity plus matter fundamentally defined by a Euclidean path integral
\be
Z[\Phi_i]=\int\prod_i[\rmd\Phi_i]\,\rme^{-S_{\rm E}[\Phi_i]},
\ee
where the fields $\Phi_i$ include Euclidean metric $g_{\mu\nu}^{\rm E}$ and $S_{\rm E}$ is the Euclidean action. This theory can suffer from a conformal instability problem \cite{Hawking:1976ja,Gibbons:1978ac,Ham09}, which we illustrate in Einstein gravity. A conformal transformation of the metric 
\be\label{WI}
g_{\mu\nu}=\Om^2 \hat g_{\mu\nu}
\ee
in the Ricci term $R$ produces a kinetic term for the conformal factor $\Om$ with the wrong sign. The Lorentzian action is
\be
\hspace{-.3cm}S=\frac{\Mpl^2}{2}\int\rmd^Dx\,\sqrt{|g|}\,R=\frac{\Mpl^2}{2}\int\rmd^D x\,\sqrt{|\hat g|}\,\Om^{D-4}\left[\Om^2\hat R+(D-1)(D-2)\hat\p_\mu\Om\hat\p^\mu\Om\right],\label{eia2}
\ee
where we have dropped boundary terms. The ghost mode $\Om$ translates into an unboundedness of the Euclidean action $S_{\rm E}\coloneqq -\rmi S$. In our conventions, $x^0=-\rmi x_D$ and $k^0=\rmi k_D$, so that, for $\hat g_{\mu\nu}=\eta_{\mu\nu}$, calling $\om\coloneqq\Om^{(D-2)/2}\Mpl\sqrt{2(D-1)/(D-2)}$,
\ba
S&=& \int\rmd^D x\,\p_\mu\om\p^\mu\om\nn
&=&\int\rmd x^0\rmd^{D-1}\bm{x}[-(\p_0\om)^2+\p_i\om\p^i\om]\nn
&=& -\rmi \int\rmd x_D\rmd^{D-1}\bm{x}[(\p_D\om)^2+\p_i\om\p^i\om]\nn
&=&-\rmi \int\rmd^D x_{\rm E}\sum_{i=1}^D(\p_i\om)^2\nn
&=& \rmi S_{\rm E}\,.\label{eia3}
\ea
Therefore, the Euclidean path integrand $\exp(\rmi S)=\exp(-S_{\rm E})$ has an anti-Gaussian term $\sim \exp[\int(\p\om)^2]$ and is unbounded from above. To avoid this arbitrarily large instability, one can deform the integration contour over the fields, in particular, taking a functional measure $[\cD g_{\mu\nu}]$ over complex Euclidean metrics (complex conformal factor), in such a way as to get a convergent result \cite{Gibbons:1978ac,Gibbons:1978ji,Christensen:1979iy,tHooft:2010xlr,tHooft:2010mvw,Marolf:2022ntb}.

The physical significance of this procedure was criticized in \cite{Schleich:1987fm,Hartle:2020glw,Mazur:1989by} and motivated another approach, where the fundamental definition of the path integral is Lorentzian and one derives (instead of assuming) the correct Euclidean path integral by integrating over a suitable contour in field space with real Lorentzian metrics \cite{Schleich:1987fm,Hartle:2020glw,Mazur:1989by,Marolf:2022ybi}.
Nonlocal quantum gravity is defined with a Lorentzian path integral, so that the issue of the conformal instability can be solved in this way. At least, there is no reason to suspect that this resolution, successful in other theories, should not work also here. As is clear from the construction at the beginning of section \ref{sec2} and in section \ref{qeasec}, all dynamics-independent tools developed to solve the conformal instability problem are available also in nonlocal quantum gravity. In particular, the York decomposition \cite{York:1973ia,York:1974psa} of metric perturbations $h_{\mu\nu}$ into transverse-traceless, scalar and gauge (coordinate) modes holds, and we also have a DeWitt metric defining the quadratic form $\langle h_{\mu\nu}, h_{\s\t}\rangle$ \cite{DeWitt:1967yk,DeWitt:1967ub,DeWitt:1967uc}. These tools allow one to handle the Jacobian factors in the path-integral measure as usual and to factorize them into Gaussian integrals over each of the metric York components. Then, since entire nonlocal form factors do not break Lorentz and diffeomorphism invariance nor hide any extra pole in field redefinitions, all the manipulations of \cite{Mazur:1989by} and of its non-perturbative extension \cite{Dasgupta:2001ue} leading to a well-defined continuation to a Euclidean path integral can be applied to the nonlocal action.

In this paper, we will not discuss the resolution of the conformal instability problem at the non-perturbative level until section \ref{sec5} because we mainly focus on perturbation theory, where the Lorentzian path integral is employed. As described in section \ref{efi}, amplitudes are calculated in Euclidean momenta and then analytically continued to imaginary Euclidean (i.e., real Lorentzian) external energies. However, \emph{at the non-perturbative level}, the Lorentzian and the Euclidean theory are distinguishable because the path integral, with its functional measure and the space of metrics on which one integrates \cite{Marolf:2022ybi,Dong:2019piw}, differs in the Euclidean and Lorentzian formulations. Redefining and analytically continuing the Lorentzian momenta of the fields appearing in the functional measure of the Lorentzian path integral produces a measure which is functionally inequivalent to the one obtained by analytically continuing the Euclidean momenta of the fields appearing in the measure of the Euclidean path integral. To put it very schematically,
\ba
\textrm{Lorentzian path integral} &\coloneqq& \int_{\textrm{metrics}\,\, (-,+,\dots,+)}\cD g_{\mu\nu}\,\rme^{\rmi S}\nn && \textrm{ with Efimov prescription on $k^0$ integration}\nn
&=& \int_{\textrm{metrics}\,\, (-,+,\dots,+)}\cD g_{\mu\nu}\,\rme^{-S_{\rm E}}\nn
&& \textrm{ with Efimov prescription on $k_D=-\rmi k^0$ integration}\nn
&=& \int_{\textrm{metrics}\,\, (+,+,\dots,+)}\cD g_{\mu\nu}^{\rm E}\,\cJ\,\rme^{-S_{\rm E}}\nn
&& \textrm{ with Efimov prescription on $k_D$ integration}\nn
&\neq& \int_{\textrm{metrics}\,\, (+,+,\dots,+)}\cD g_{\mu\nu}^{\rm E}\,\rme^{-S_{\rm E}}\nn
&& \textrm{ with naive analytic continuation}\nn
&=& \textrm{Euclidean path integral with naive analytic continuation},\nn\label{scheme}
\ea
where the action is expressed as an integral in momentum space and the Jacobian $\cJ=\cD g_{\mu\nu}/\cD g_{\s\t}^{\rm E}$ can be inferred from \cite{Dasgupta:2001ue}. In particular, nonlocal quantum gravity is defined by the left-hand side of the first line of \Eq{scheme}, not by the last line.

From this discussion, it stems that the conformal instability problem does not affect perturbative results because, in general, the Lorentzian perturbation theory is well defined \cite{Ham09,Ordonez:1985kz}. Let us expand on this statement. In order for the Lorentzian path integral to be well defined in perturbation theory, the tree-level graviton kinetic term in Feynman prescription $k^2\to k^2-\rmi\e$ must generate a Gaussian term in the path integral that makes it convergent. If this was the case, then we would have convergence order by order in perturbation theory because, at higher orders, one only has to include vertex insertions, which do not spoil the tree-level convergence property. This is because all correlation functions in perturbation theory are written as an expansion of the interactions  $S_{\rm int}$: schematically for a generic field $\Phi$, 
\be
\langle \Phi(x_1) \dots \Phi(x_n)\rangle=\sum_n\int \Phi(x_1) \dots \Phi(x_n)\, \rmi^n S_{\rm int}^n\,\frac{\rme^{\rmi S_0}}{n!}\,,
\ee
where $S_0=\int\Phi\cO\Phi$ and $\cO = \cO^{\rm K} + \cO^{\rm gf}$ is the tree-level kinetic term (inverse of the propagator), where $\cO^{\rm K}$ is the kinetic term coming from the bare action and $\cO^{\rm gf}$ is a contribution coming from the gauge-fixing action. In order to explicitly compute any perturbative amplitude, we should commute the functional integral with the sum. This issue is related to the Borel summability of the perturbative series that deserves to be investigated also in gravity and, in particular, in a finite theory of quantum gravity in the QFT framework.

If one naively considered only the gauge-invariant part $\cO^{-1}_{\rm K}$ of the propagator, the convergence problem would persist after the $k^2\to k^2-\rmi\e$ prescription because the spin-2 and spin-0 modes in $\cO^{\rm K}$ have opposite sign, so that the spin-0 mode would generate an anti-Gaussian profile. Thus, the role of $\cO^{\rm gf}$ cannot be ignored. If, after a Weyl transformation \Eq{WI}, a suitable gauge choice exists such that the eigenvalues of the kinetic operator of the conformal factor $\Om$ are all zero or negative in Euclidean signature, then there is convergence of the Euclidean path integral at the perturbative level. 

To show this, it is sufficient to consider the gravitational sector of nonlocal quantum gravity and the generic action \Eq{GA} where now $\g_0$, $\g_2$ and $\g_4$ are generic form factors which vanish for Einstein gravity and are constant in Stelle gravity, and $S_{\rm gf}$ is the gauge-fixing action. In a Weyl invariant theory, the metric $g_{\mu\nu}$ composing the measure and the curvature tensors is the one before making explicit the dilaton dependence. Decompose the metric as in \Eq{gh} and derive the perturbed action for the dimensionless spin-2 field $h_{\mu\nu}$ (the graviton polarization modes are inside this object). On Minkowski background, the kinetic term is $\propto h^{\mu\nu}\cO_{\mu\nu\s\t}h^{\s\t}$, where the operator $\cO_{\mu\nu\s\t}$ is given in section \ref{sec3} and is a generalization of the higher-derivative expression found in \cite{Accioly:2002tz}.

Consider now a Weyl transformation producing the incriminated kinetic term with the wrong sign for the conformal factor $\Om$. To isolate the troublesome scalar mode, it is sufficient to pick a conformally flat metric $\hat g_{\mu\nu}=\eta_{\mu\nu}$ \cite{Hartle:2020glw,Mazur:1989by,Hawking:1978jz,Komargodski:2011vj}:
\be
g_{\mu\nu} = \Om^2\, \eta_{\mu\nu}\quad\Longleftrightarrow\quad h_{\mu\nu}=(\Om^2-1)\eta_{\mu\nu}\eqqcolon \phi\eta_{\mu\nu}\,,
\ee
where $\phi$ is a scalar. Therefore,
\ba
h^{\mu\nu}\cO_{\mu\nu\s\t}h^{\s\t} &=& \phi\eta^{\mu\nu}\eta^{\s\t}\cO_{\mu\nu\s\t}\phi\nn
&=&\phi[(D-1)c_0+\bar{c}_0+2(D-1)\bar{\bar{c}}_0]\phi\nn
&\eqqcolon&\phi\cK\phi\,,
\ea
where we used \Eqq{E23}, the Barnes--Rivers projectors \Eq{projectors} and the ensuing properties
\bs\ba
&&\eta^{\mu\nu}\eta^{\s\t}P^{(1)}_{\mu\nu\s\t}=0=\eta^{\mu\nu}\eta^{\s\t}P^{(2)}_{\mu\nu\s\t}\,,\\ &&\eta^{\mu\nu}\eta^{\s\t}P^{(0)}_{\mu\nu\s\t}=D-1\,,\\ &&\eta^{\mu\nu}\eta^{\s\t}\bar{P}^{(0)}_{\mu\nu\s\t}=1\,,\\ &&\eta^{\mu\nu}\eta^{\s\t}\bar{\bar P}_{\mu\nu\s\t}^{(0)}=2(D-1)\,.
\ea\es
Intuitively, only the spin-0 projectors survive because they act on a scalar. From the expressions \Eq{E25} for the coefficients $c_0$, $\bar{c}_0$ and $\bar{\bar{c}}_0$, in momentum space we have
\ba
\cK&\!=\!&\frac{\Mpl^{2}}{4} k^2\big\{(D-1)(D-2)-2(D\la-1)^2\la_1\nn
&&+k^2\left[4(D-1)^2\g_0'+D(D-1)\g_2'+(D\la-1)^2\la_2\g_2'\right]\big\},
\ea
where $\la$, $\la_1$, and $\la_2$ are gauge parameters (or, more generally, functions of $k^2$) appearing in the gauge-fixing contribution $\cO^{\rm gf}$.

If we ignored the gauge-fixing terms and set $\la=\la_1=\la_2=0$, then $\cK>0$ in Euclidean signature for non-negative form factors $\g_{0,2}'\geq 0$ and the Euclidean action $S_{\rm E}\sim-\int\phi\cK\phi$ would be unbounded from above. This is the conformal instability problem. However, by virtue of gauge invariance, if we can find a range of gauge choices where $\cK\leq 0$ for all momenta $k$, then we can conclude that such problem is a gauge artefact and that the scalar mode $\phi$ does not propagate ($\cK=0$) or that it propagates with a kinetic term of the ``straight'' sign ($\cK<0$). In general, for $\g_2'\neq 0$ this happens when
\ba
&&\la_1\geq\frac{(D-1)(D-2)}{2(D\la-1)^2}\,,\label{gaueg}\\
&&\la_2\leq-\frac{D-1}{(D\la-1)^2}\left[4(D-1)\frac{\g_0'}{\g_2'}+D\right].\label{gaust}
\ea
The first of these conditions appears in all theories with an Einstein--Hilbert term and implies that both $\la$ and $\la_1$ can be chosen to be constant. In $D=4$ dimensions, $\la_1\geq 3/(4\la-1)^2$ and infinitely many gauges with $\la\neq 1/4$ can fulfill \Eqq{gaueg}, including Julve--Tonin ($\la=0$) and de Donder ($\la=1/2$, $\la_2=0=\la_3$), but not the Feynman gauge ($\la=1/2$, $\la_1=1$, $\la_2=0=\la_3$).

We give three applications of the inequalities \Eq{gaueg} and \Eq{gaust}:
\begin{itemize}
\item \emph{Einstein gravity}: $\g_0=\g_2=\g_4=0$. The gauge choice making the kinetic term vanish or with negative eigenvalues is only \Eq{gaueg}.
\item \emph{Stelle gravity}: $\g_0,\g_2,\g_4$ constant. Then, we impose \Eqqs{gaueg} and \Eq{gaust}. This range includes the Julve--Tonin gauge and, depending on the ratio $\g_0'/\g_2'$, also the de Donder gauge.
\item \emph{Nnlocal quantum gravity}: $\g_0,\g_2,\g_4$ asymptotically polynomial form factors, where $\g_4=0$ in the nonminimally coupled version of the theory. The ratio in \Eq{gaust} is
\be
\frac{\g_0'}{\g_2'}=- \frac{1}{4(D-1)}\left[(D-2)\frac{\rme^{\H_0}-1}{\rme^{\H_2}-1} + D\right],
\ee
so that
\be
\la_2\leq \bar \la_2\coloneqq \frac{(D-1)(D-2)}{(D\la-1)^2}\frac{\rme^{\H_0}-1}{\rme^{\H_2}-1}.\label{gaust2}
\ee
When $\H_0=\H_2$, one has $\g_0'/\g_2'=-1/2$ and \Eq{gaust2} becomes
\be\label{gaunl}
\la_2\leq\frac{(D-1)(D-2)}{(D\la-1)^2},
\ee
which forms a system with \Eqq{gaueg}. In $D=4$ dimensions, $\la_2\leq 6/(4\la-1)^2$ and both the Julve--Tonin and de Donder gauge are included. When $\H_0\neq\H_2$, it is easy to check that the function $\bar \la_2$ is bounded from above and from below for all $k^2$, so that it is sufficient to take its lower limit as the upper bound for $\la_2$.\footnote{From the asymptotic limits discussed below \Eq{MCFormFactor}, it turns out that both in the IR and in the UV $[\rme^{\H_0(z)}-1]/[\rme^{\H_2(z)}-1]\simeq p_0(z)/p_2(z)$. For monomials $p_{0,2}(z)=a_{0,2}z^n$, this asymptotic limit corresponds to the upper (respectively, lower) bound of $\bar \la_2(z)$ if $a_0<a_2$ (respectively, $a_0>a_2$). The lower (respectively, upper) bound is given by the other two local extrema of $\bar \la_2(z)$.} Therefore, $\la_2$ can be taken to be constant also in this case.
\end{itemize}


\section{Discussion}\label{sec5}

In this paper, we have formalized the path integral of nonlocal quantum gravity, in particular, its version with asymptotically polynomial operators. We have presented the tree-level propagator in arbitrary gauge and discussed how the conformal instability problem of the Euclidean version of the theory disappears in perturbation theory thanks to gauge invariance.

The procedure detailed in section \ref{secpi} works only at the perturbative level and it differs from the one of \cite{Mazur:1989by} by the explicit use of the gauge-fixing action. This method has the advantage of drastically simplifying the derivation of the main result. Since the tree-level propagator is calculated anyway as one of the building blocks of Feynman diagrams, this approach is fairly parsimonious in computation time.

To date, and modulo some very preliminary exceptions \cite{Ghoshal:2022mnj}, 
 nonlocal quantum gravity has been studied only with perturbative techniques in relation with scattering amplitudes, perturbative unitarity, renormalization, black-hole solutions and cosmology. However, the conformal instability problem could arise when considering non-perturbative processes and the full path integral is not expanded in terms of the interactions.
As in the perturbative case, once the Euclidean path integral is derived from the Lorentzian one, a non-trivial choice of the gravitational measure of the path integral can make it convergent. The physical interpretation is that this measure would enhance the weight of strong metric fluctuations pushing away physical fluctuations from the conformal unbounded abyss \cite{Ham09}. This seems indeed to be the case, both on a lattice at strong coupling \cite{Hamber:1984kx,Berg:1985ka} and from a non-perturbative extension \cite{Dasgupta:2001ue,Ambjorn:2002gr} of the perturbative calculation of  \cite{Mazur:1989by}. In the case of the Einstein--Hilbert action, after a Wick rotation of the non-perturbative Lorentzian path integral defined on a causal dynamical triangulation (CDT), it turns out that the term responsible for the conformal instability is cancelled by a compensating term arising from the integration of the Faddeev--Popov determinant that arises in the path-integral measure when gauge fixing \cite{Dasgupta:2001ue,Ambjorn:2002gr}. A cancellation mechanism virtually identical to this should apply also in nonlocal gravity, since it amounts to a non-perturbative upgrade of the argument exposed in section \ref{secpi} based on the same physical principle of gauge invariance. In particular, the calculation of the measure and the cancellation mechanism in \cite[section 4]{Dasgupta:2001ue} apply \emph{verbatim} also to nonlocal quantum gravity because the conformal divergence stems from the $R$ term in the action, \Eqqs{eia2} and \Eq{eia3}; in fact, a conformal transformation of $O(R^2,R_{\mu\nu}R^{\mu\nu})$ operators with or without nonlocal form factors does not produce terms with the opposite sign. Thus, keeping the leading conformal divergence from the Einstein--Hilbert term $R$ and integrating over the conformal mode after a field redefinition introduced in \cite{Distler:1988jt}, the compensation of the unbounded term by the functional determinant takes place.
Numerical analyses in CDT support this conclusion for Einstein gravity. Upon discretization, the action is bounded from below for a fixed discrete spacetime volume. The minimum of the action happens when the ratio $N_{22}/N_3$ between the number $N_{22}$ of so-called 2-2 tetrahedra and the total number $N_3$ of tetrahedra is minimal. Numerical simulations in three dimensions show that the expectation value $\langle N_{22}/N_3\rangle$ stays positive and away from zero in the continuum limit \cite{Ambjorn:2000dja,Ambjorn:2001br}, meaning that the kinetic term of the conformal mode never dominates the dynamics and its contribution to the path integral is suppressed in the continuum Lorentzian limit \cite{Dasgupta:2001ue,Ambjorn:2002gr}. All of this is relevant also for nonlocal quantum gravity, since CDT is not an independent proposal of quantum gravity but, rather, a regularization method that can be applied to any path integral with a diffeomorphism-invariant action. Entire form factors do not add any extra poles nor change the sign of kinetic terms and asymptotically local quantum gravity on a CDT should inherit the same qualitative properties found for Einstein gravity. These theoretical and numerical aspects will deserve a check in the future in the nonlocal case.


\section*{Acknowledgments}

The authors are supported by grant PID2020-118159GB-C41 funded by MCIN/AEI/10.13039/ 501100011033. G.C.\ thanks F.\ Briscese and R.\ Loll for useful discussions.


\end{document}